\documentclass[aps,prl,superscriptaddress,twocolumn]{revtex4}
\usepackage{graphicx}
\usepackage{amsmath}
\usepackage{xcolor}
\usepackage{epstopdf}

\begin{document}
\title{Niobium diselenide superconducting photodetectors}
\author{G. J. Orchin}
\affiliation{School of Engineering, University of Glasgow, Glasgow  G12 8QQ, UK}
\author{D. De Fazio }
\affiliation{Cambridge Graphene Centre, University of Cambridge, Cambridge CB3 0FA, UK}
\author{A. Di Bernardo }
\affiliation{Department of Materials Science \& Metallurgy, University of Cambridge, Cambridge CB3 0FS, UK}
\author{M. Hamer}
\affiliation{National Graphene Institute, The University of Manchester, Manchester M13 9PL, UK}
\author{D. Yoon}
\affiliation{Cambridge Graphene Centre, University of Cambridge, Cambridge CB3 0FA, UK}
\author{A. R. Cadore}
\affiliation{Cambridge Graphene Centre, University of Cambridge, Cambridge CB3 0FA, UK}
\author{I. Goykhman}
\affiliation{Cambridge Graphene Centre, University of Cambridge, Cambridge CB3 0FA, UK}
\author{K. Watanabe}
\affiliation{National Institute for Materials Science, 1-1 Namiki, Tsukuba 306-0044, Japan}
\author{T. Taniguchi}
\affiliation{National Institute for Materials Science, 1-1 Namiki, Tsukuba 306-0044, Japan}
\author{J. W. A. Robinson}
\affiliation{Department of Materials Science \& Metallurgy, University of Cambridge, Cambridge CB3 0FS, UK}
\author{R. V. Gorbachev}
\affiliation{National Graphene Institute, The University of Manchester, Manchester M13 9PL, UK}
\author{A. C. Ferrari}
\affiliation{Cambridge Graphene Centre, University of Cambridge, Cambridge CB3 0FA, UK}
\author{R. H. Hadfield}
\affiliation{School of Engineering, University of Glasgow, Glasgow  G12 8QQ, UK}
\begin{abstract}
We report the photoresponse of niobium diselenide (NbSe$_2$), a transition metal dichalcogenide (TMD) which exhibits superconducting properties down to a single layer. Devices are built by using micro-mechanically cleaved 2 to 10 layers and tested under current bias using nano-optical mapping in the 350mK-5K range, where they are found to be superconducting. The superconducting state can be broken by absorption of light, resulting in a voltage signal when the devices are current biased. The response found to be energy dependent making the devices useful for applications requiring energy resolution, such as bolometry, spectroscopy and infrared imaging.
\end{abstract}
\maketitle
The potential of superconducting materials for low noise photon detection has long been recognised\cite{Richards1994, Kadin1990}. Several concepts exploiting superconductors have further evolved, including transition edge sensors (TES)\cite{Irwin1996}, Kinetic Inductance Detectors (KID)\cite{Baselmans2012} and Superconducting Nanowire Single Photon Detectors (SNSPD)\cite{Goltsman2001}. These devices have found applications in quantum optics\cite{Hadfield2009}, quantum key distribution\cite{Takesue2007}, single photon LIDAR\cite{Mccarthy2013} and infrared astronomy\cite{Richards1994}. Superconducting Nanowire Single Photon Detectors (SNSPDs) can be built with$>$90\% single photon detection efficiency at 1550nm\cite{Marsili2013}. In these detectors a light pulse impinges on a current-biased superconducting nanowire, generating a local resistive region (hotspot)\cite{Natarajan2012}. This makes the superconducting current diverge around it, increasing current density in the surrounding areas\cite{Natarajan2012}. The devices can be operated under bias conditions so that the carrier density in the regions adjacent to the hotspot breaks the superconducting state, causing the formation of a resistive barrier\cite{Natarajan2012}. Current is then forced to cross the barrier, which can be sensed by a readout circuit\cite{Natarajan2012}. One route to improving the device photoresponse at the telecommunication wavelength 1550nm is to scale down the nanowire width\cite{Lusche2014}. This provides heat confinement which, in turn, facilitates the formation of the resistive barrier\cite{Lusche2014}. However, this requires nanometre-precision in lithographic tools\cite{Korneeva2018}. An alternative path could be decreasing the device thickness.

NbSe$_2$ is a superconductor with a bulk superconducting transition temperature ($T_c$)$\sim$7.2K\cite{RevoJPCS1965}. Crystals can be exfoliated to produce flakes down to one layer (1L)\cite{FrinPRL1972, Cao2015}. Ref.\citenum{Cao2015} showed that NbSe$_2$ is superconducting down to 1L, with $T_c\sim$2K, while 2L-NbSe$_2$ have $T_c\sim$5K\cite{Cao2015}, above the 4.2K liquid He temperature used in SNSPDs measurements setups\cite{Hadfield2009}. This paves the way to the study of thin superconducting detectors\cite{Lusche2014}.

Here we use micron-sized 2 to 10L NbSe$_2$ flakes as superconducting detectors for 1550nm pulsed illumination in the 5K-300mK range. We find a non-linear response with a minimum noise equivalent power (NEP)$\sim$40fW Hz$^{-0.5}$, at a laser pulse energy of 27fJ. This is comparable with cryogenic graphene hot electron bolometers\cite{Yan2012} and commercial InGaAs photodiodes\cite{TL,Donati}.
\begin{figure*}
\includegraphics[width=160mm]{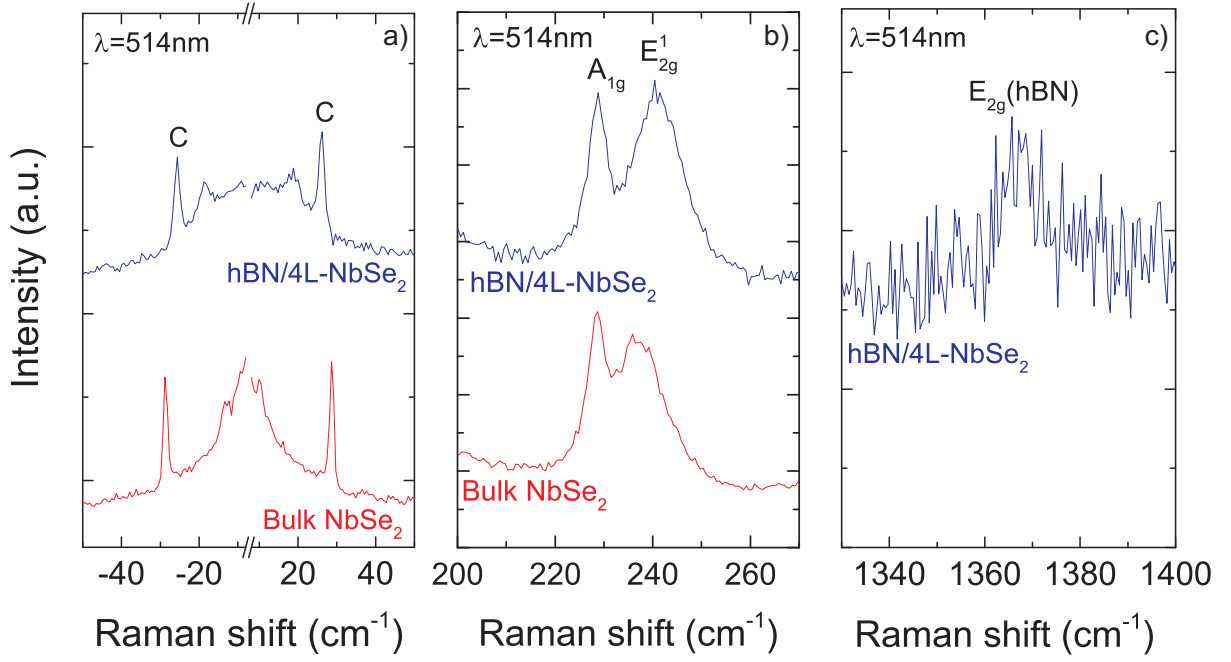}
\caption{a-c) Raman spectra of hBN/4L-NbSe$_2$ (blue) and bulk NbSe$_2$ (red)}
\label{Raman}
\end{figure*}

The detectors are fabricated as follows. We use a substrate with 285nm SiO$_2$ on Si, for optimal optical contrast\cite{CasiNL2007,BenaN2011}. On this we fabricate Ti/Au (5nm/50nm) contacts by evaporation and lift-off. Bulk NbSe$_2$ is sourced from HQ Graphene\cite{HQG}. Ref.\citenum{ElbaSST2013} showed that NbSe$_2$ can degrade in air due to reaction of Nb with O$_2$. Thus, an exfoliated sample left unprotected for several hours would be unlikely to show superconductivity\cite{Cao2015}. It is therefore necessary to encapsulate NbSe$_2$ with an impermeable membrane, such as a flake of hexagonal boron nitride (hBN)\cite{Cao2015}. Here we use hBN crystals grown by the temperature-gradient method under high pressure and high temperature\cite{TaniJCG2007}. hBN and NbSe$_2$ flakes are exfoliated by micromechanical cleavage on Nitto Denko tape\cite{NovoPNAS2005}, then exfoliated again on a polydimethylsiloxane (PDMS) stamp, placed on top of a glass slide, for inspection under optical microscope. This yields large (up to$\sim$100$\mu$m) flakes\cite{Cast2DM2014}. Few-layer (FL) flakes are selected based on optical contrast and then stamped on the contacts with a micro-manipulator. FL-hBN flakes are exfoliated on tape, then on SiO$_2$/Si for microscope inspection. Thin flakes (up to 10nm), selected by optical contrast, are picked with a polycarbonate (PC)/PDMS stack prepared on a microscope glass slide\cite{PalaNC2016}. The stack is then mounted on a micromanipulator under an optical microscope and transferred on FL-NbSe$_2$ to protect it from air-exposure. The temperature is raised to$\sim$100$^{\circ}$C to release PC, which is then dissolved in chloroform\cite{PurdNC2018}. 30 devices of 2-10L-NbSe$_2$ were prepared to check reproducibility and the effect of thickness on device performance. We did not use 1L-NbSe$_2$ due to its $T_c$ being lower than 4.2K\cite{Cao2015}.

Raman spectroscopy in vacuum is performed by mounting a Oxford Instruments vacuum stage under a 50x ultra-long working distance objective of an Horiba LabRam Evolution spectrometer. A turbomolecular pump is used to achieve a base pressure$\sim$10$^{-5}$Torr. Measurements are done with 514nm excitation wavelength, with a 1800grooves/mm grating and a spectral resolution$\sim$0.3cm$^{-1}$. The power is kept below 300$\mu$W to avoid any damage.

Fig.\ref{Raman}a plots the low frequency ($<100$cm$^{-1}$) spectrum of a NbSe$_2$ flake after hBN encapsulation and that of bulk NbSe$_2$. The shear modes (C) are due to the relative motions of the atomic planes\cite{FerrNN2013,ZhanPRB2013,TanNM2012}. The position of the C peak, Pos(C), can be used to derive the number of layers (N) as\cite{TanNM2012}:
\begin{equation}
\label{Eq1}
N=\frac{\pi}{2\cos^{-1}\left[\frac{\mathrm{Pos(C)_N}}{\mathrm{Pos(C)_{\infty}}}\right]}
\end{equation}
where $\mathrm{Pos(C)_{\infty}}$ is Pos(C) in the bulk. Fig.\ref{Eq1}a gives $\mathrm{Pos(C)_{\infty}}\sim$28.6cm$^{-1}$ (red curve) and$\sim$26.2cm$^{-1}$ (blue curve). Thus, N$\sim$4. The hBN E$_{2g}$ peak\cite{ReicPRB2005,ArenNL2006} is at $\sim$1367cm$^{-1}$ in Fig.\ref{Eq1}c, and it has FWHM$\sim$9cm$^{-1}$, as expected for$\sim$10nm hBN\cite{NemaPRB1981}.

Optical measurements are then performed in a closed cycle 2-stage Pulse Tube with an additional $^3$He stage capable of reaching a base temperature$\sim$350mK, equipped with a bespoke confocal microscope with piezoelectric positioners\cite{Li2016}. This allows repositioning of the focussed laser spot without sample warm up.
\begin{figure}
\centerline{\includegraphics[width=80mm]{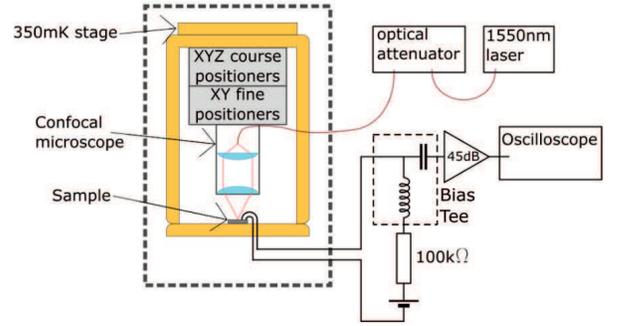}}
\caption{Experimental setup used to measure the photoresponse. Attenuated laser pulses are focussed on a current biased sample using a cryogenic confocal microscope. The resulting voltage pulses are amplified at RT before being counted by a universal counter or viewed on an oscilloscope.}
\label{setup}
\end{figure}
\begin{figure}
\includegraphics[width=80mm]{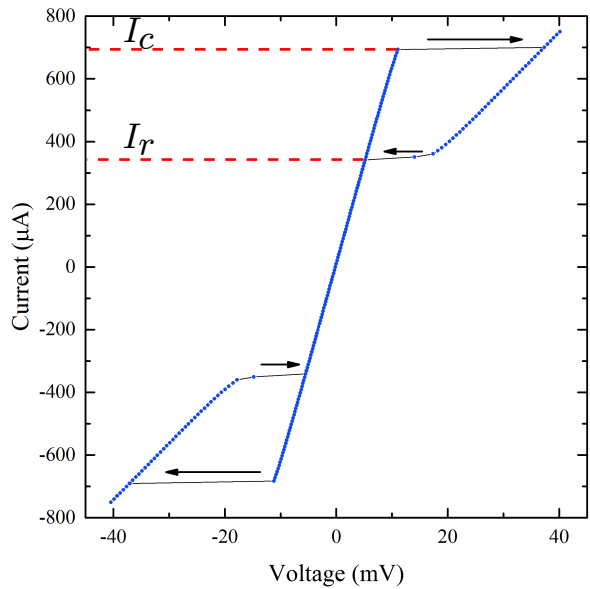}
\caption{Current-voltage (I-V) measurement of a 4L-NbSe2 device at 0.36K.}
\label{critical_current}
\end{figure}
\begin{figure*}
\centerline{\includegraphics[width=160mm]{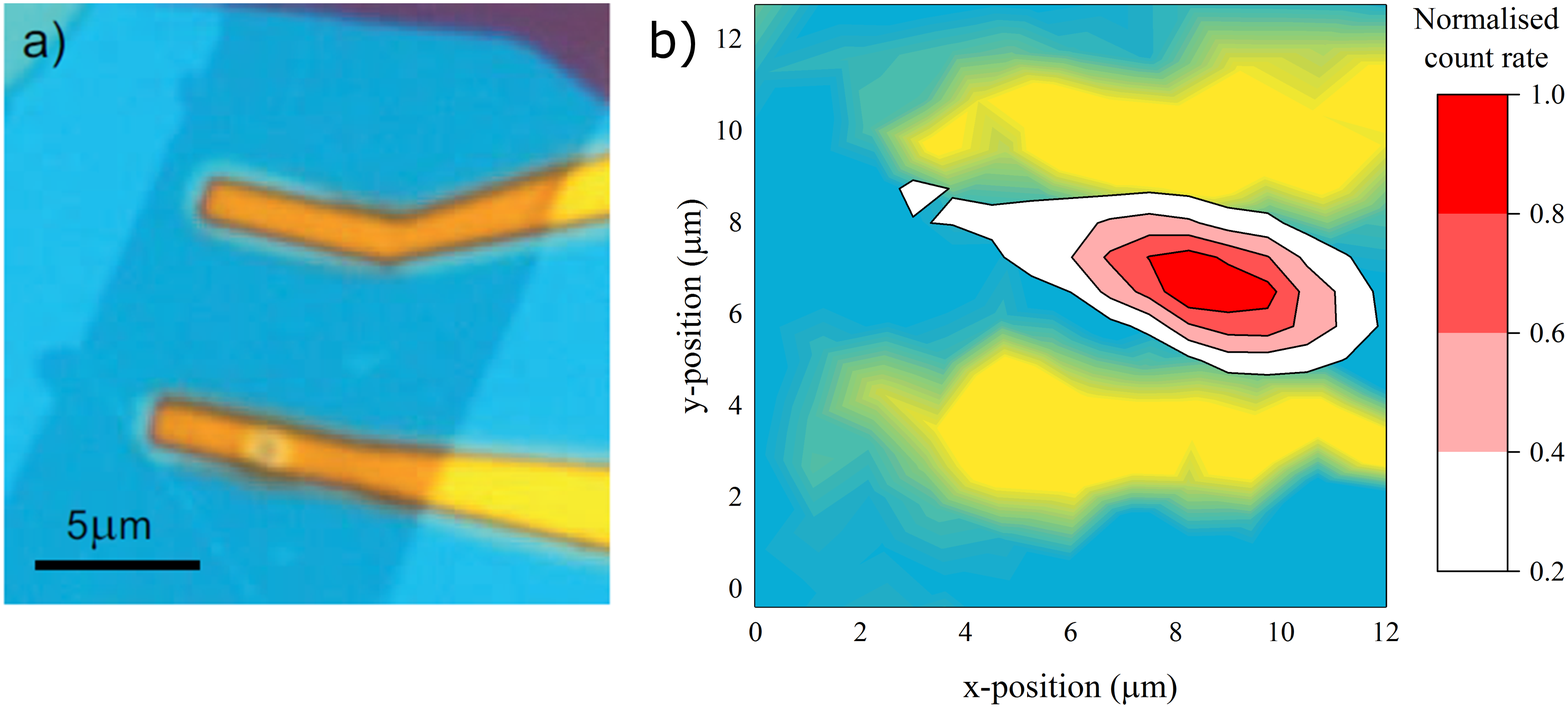}}
\caption{a) Optical image of 4L-NbSe$_2$ flake, covered by$\sim$10nm hBN (light blue) transferred on Au electrodes. b) False color image of the same sample at 5K using 1550nm illumination. White to red contour lines show the normalized count rate of 17fJ laser pulses, thus the optically sensitive area of the detector.}
\label{map}
\end{figure*}

The current necessary to break the superconducting state, the critical current $I_c$\cite{Tinkham2004Introduction}, is measured at the cryostat base temperature (T), Fig.\ref{critical_current}a. $I_c$ is identified by the sudden change of slope in the I/V characteristics. We find I/V curves to be hysteretic below 5K. Indeed, in the reverse sweep, the current at which superconductivity is re-established, the re-trapping current, $I_{r}$\cite{Zotova2013}, is as low as half $I_c$. This is because electrical heating generated in the resistive state cannot be dissipated, and increases the sample $T$. We obtain $I_c(0)=677\mu A$ and $T_c=6K$.

Using a 1$\mu$W CW 1550nm laser, the confocal microscope is raster scanned over the device using piezoelectric positioners. The resulting reflection is used to construct an image of the device area at 5K (Fig.\ref{map}b, blue-yellow). To locate the photosensitive area, the NbSe$_2$ devices are current biased at 0.9I$_c$ (227$\mu$A) using the circuit shown in Fig.\ref{setup}. A programmable pulse generator is then used to drive a 1550nm laser diode at 1MHz repetition rate with a 20ns pulse width. The laser is attenuated to low power ($\sim$17fJ per pulse) and scanned over the device. The resulting voltage signals are amplified with 45dB gain over the bandwidth 10kHz-500MHz using a room temperature (RT) low noise amplifier (RF Bay LNA-545) and voltage pulses are counted by a universal counter (Agilent 53132A). As shown in Fig.\ref{map}b (white-red contour lines), the photosensitive area is centered between the two electrical contacts indicating that the superconducting NbSe$_2$ is the photosensitive element.
\begin{figure}
\centerline{\includegraphics[width=90mm]{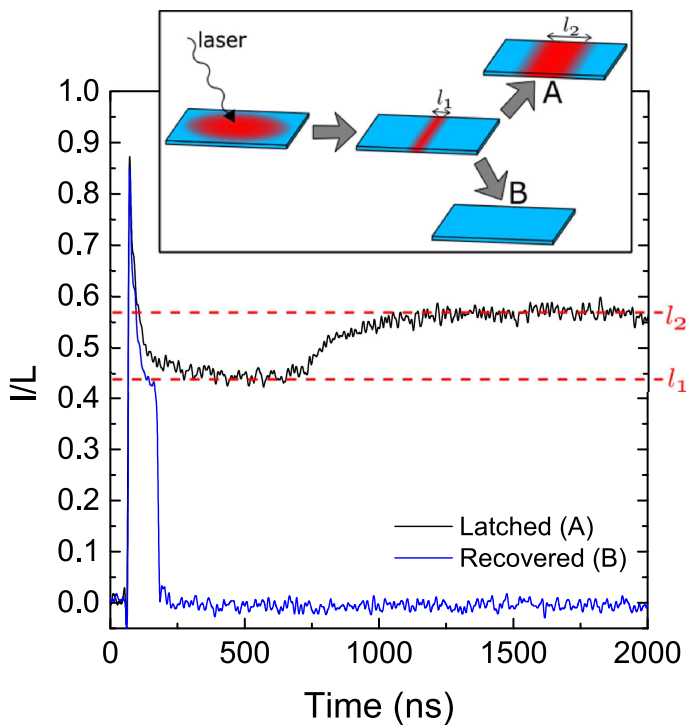}}
\caption{Oscilloscope traces showing the critical resistive length phenomenon around $l_1$. The inset illustrates the two possible hotspot evolutions: (A) growth to $l_2$ and (B) decay to 0 with recovery of superconductivity.}
\label{Pulses_around_l1}
\end{figure}

Using the readout circuit in Fig.\ref{setup}a a 4L-NbSe$_2$ flake is biased at 390$\mu$A, just above the measured $I_{r}$, Fig.\ref{critical_current}. It is illuminated with single shot laser pulses and the resulting voltage signals are observed using an oscilloscope. At this current the detector is driven to a persistent resistive state (latched) by a strong optical pulse but recovers superconductivity if a low power pulse is used. When the laser pulses are attenuated to$\sim$500fJ per pulse, the device sometimes recovers superconductivity and other times latches to a steady state voltage. Examples of these pulse shapes are in Fig.\ref{Pulses_around_l1}, where the voltage pulses are normalized to the voltage corresponding to the full length of the NbSe$_2$ flake, $L$. In both traces, the length of the resistive region initially settles to $0.44L$, where they then diverge. The latching pulse (A) grows to $0.56L$ and the recovered pulse (B) drops to 0 in $\sim$20ns, returning to the superconducting state.

In order to describe the process by which the absorption of radiation causes an electrical response from NbSe$_2$, we consider how an optically stimulated resistive region would evolve within a superconducting flake. This is governed by the balance of Joule heating and heat dissipation\cite{Skocpol1974, Blois2017}. Assuming uniform thermal conductivity, heat capacity and current density ($J = I/A$ where $A$ is the cross-sectional area of the flake), the T distribution in the flake will tend towards being one dimensional due to heat diffusion\cite{Zotova2012}.

We model the device as initially at a constant temperature ($T_{bath}$), in thermal equilibrium with its environment. Current is passed by the Ti/Au electrodes. Their thickness is 55nm, at least 10 times larger than our NbSe$_2$. We thus assume that the contacts maintain each end of the device at $T_{bath}$. This geometry was studied analytically in Ref.\cite{Skocpol1974} for the time independent case, when Joule heating is balanced by thermal dissipation. The current density ($J$) required to sustain a resistive region of length $l$, when the material properties are T independent, can be written as\cite{Skocpol1974}:
\begin{equation} J^2 = \frac{2\alpha \Big(T_c - T_{bath}\Big)cosh\Big(\frac{L}{2}\sqrt{\frac{\alpha}{\kappa d}}\Big)}{\rho d \Big\{cosh\Big(\frac{L}{2}\sqrt{\frac{\alpha}{\kappa d}}\Big) - cosh\Big[\Big(\frac{L}{2} - l\Big)\sqrt{\frac{\alpha}{\kappa d}}\Big]\Big\}}
\label{Ireq}
\end{equation}
Where $\kappa$ is the T dependent thermal conductivity of NbSe$_2$\cite{Boaknin2003}, $l$ is the length of the self-sustaining resistive region, $\rho$ is the resistivity of the flake in the normal state, $d$ is the thickness of the superconducting flake and $\alpha$ is the heat transfer coefficient to the substrate, a sample dependent parameter.

Eq.\ref{Ireq} shows that $J$ required to sustain a resistive region of length $l$ reaches a minimum, $J_{latch}$, when $l=L/2$. Therefore, the bias current must be $>I_{latch}=AJ_{latch}$ for the superconducting flake to have a steady state resistance.

Using Eq.\ref{Ireq} to solve for $l$, we find that any current $>I_{latch}$ has two solutions $l_1$ and $l_2=L-l_1$, with $l_1$ the smaller of the two lengths. For these, the electrical power dissipated in the resistive region will be balanced by thermal diffusion. However, although both are valid solutions, only $l_2$ is stable in time. $l_1$ is unstable since if $l<l_1$, the resistive region will cool and therefore shrink, whereas, if $l>l_1$, the region will be heated and therefore expand. In either case the resistive length will be driven away from $l_1$.
\begin{figure}
\centerline{\includegraphics[width=90mm]{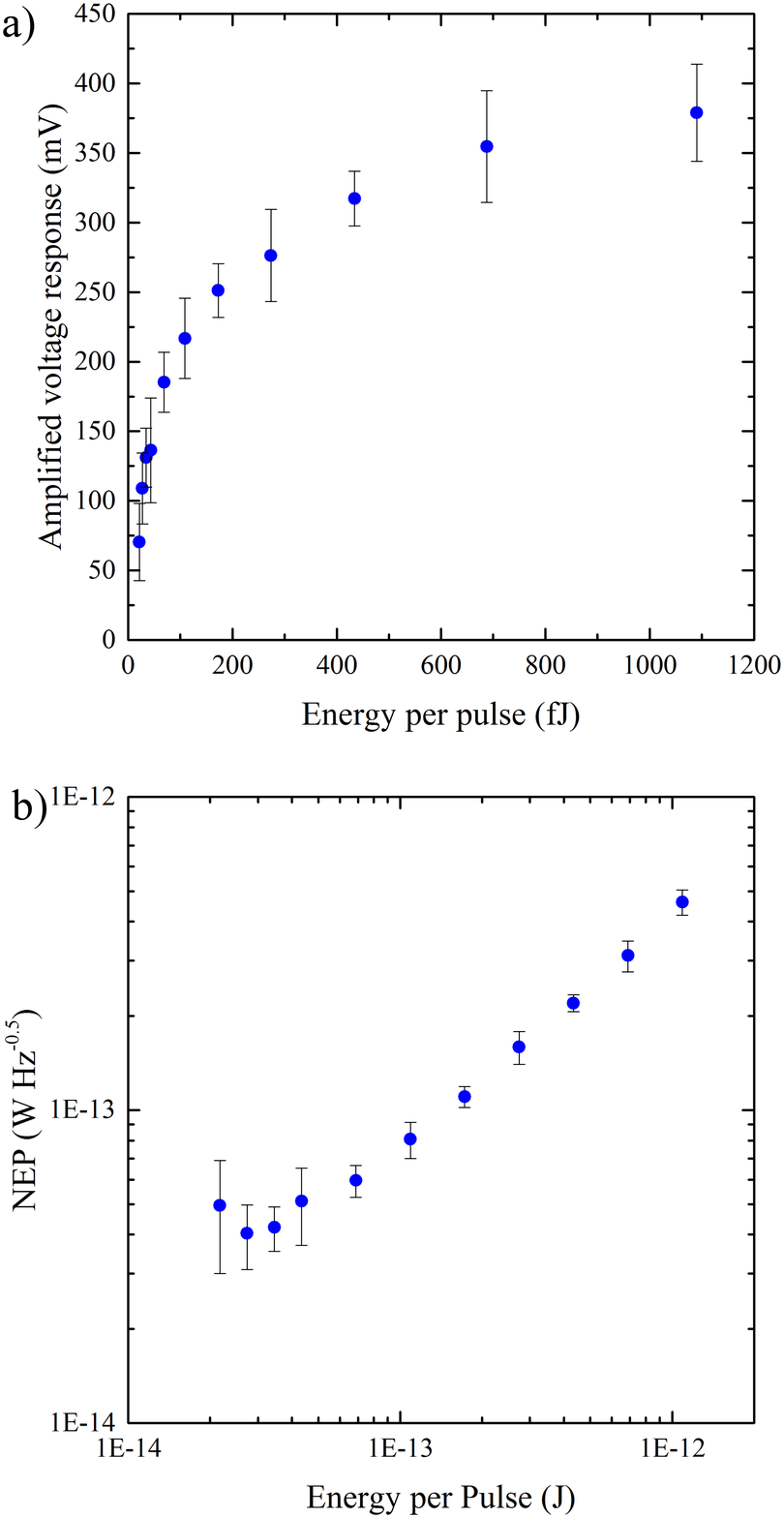}}
\caption{a) Measured voltage pulse as a function of laser energy for 230$\mu A$ bias current at 5K. b) NEP as a function of laser energy, calculated by dividing the spectral noise density of the system by the responsivity of the detector.}
\label{height_vs_atten}
\end{figure}

In contrast, there will be negative electro-thermal feedback around $l_2$, thus a self sustaining resistive region will settle to this length. $l_1$ can therefore be viewed as a critical length. If the hotspot length decays below $l_1$, the resistive region will not be self-sustaining and superconductivity will be re-established, Fig.\ref{Pulses_around_l1} pulse B. If the hotspot length remains above $l_1$, it will eventually grow to $l_2$. This results in a latched resistive state, with the steady state resistance of the device proportional to $l_2$, Fig.\ref{Pulses_around_l1} pulse A.

To demonstrate the energy dependent nature of the photoresponse, the NbSe$_2$ flake in Fig.\ref{map} is biased at 230$\mu$A (0.91 I$_c$) at 5K and illuminated with single shot laser pulses of energy ranging from 20 to 1100fJ, Fig.\ref{height_vs_atten}a. The voltage response of the device, Fig.\ref{height_vs_atten}a, increases with laser energy and starts saturating at$\sim$500fJ. This behavior can be understood as the laser driving a section of the flake into the resistive state. As the laser energy increases, so does the size of the resistive region. This occurs until the resistive region approaches the total length of the flake, causing saturation of the voltage response. The NEP of the detector is calculated by dividing the noise spectral density in the signal line by the responsivity of the detector\cite{Donati}. As shown in Fig.\ref{height_vs_atten}b, the NEP reaches a minimum$\sim$40fW Hz$^{-0.5}$. For comparison, infrared graphene hot electron bolometers have an electrical NEP$\sim$33fW Hz$^{-0.5}$ at 5K\cite{Yan2012} and commercial InGaAs photodiodes 5-1000fW Hz$^{-0.5}$\cite{TL,Donati}.

In summary, we prepared hBN encapsulated NbSe$_2$ flakes ranging from 2-10 layers. These are superconducting and photosensitive at 1550nm, with a power dependent response when illuminated. The optical NEP of the detectors is as low as 40fW Hz$^{-0.5}$ at 5K, competitive with graphene based hot electron bolometers and commercial photodiodes. The optical absorption could be further improved by integration with optical cavities, waveguides or plasmonic structures.

We thank L. Baker, D. Morozov and R. Heath for assistance with cryogenic measurements. We acknowledge funding from ERC grants IRIS 649604, Hetero2D, EPSRC Grants EP/K01711X/1, EP/K017144/1, EP/N010345/1, EP/L016087/1, and the EU Graphene and Quantum Flagships and Elemental Strategy Initiative conducted by the MEXT, Japan and the CREST (JPMJCR15F3), JST.


\begin{thebibliography}{100}

\bibitem{Richards1994} P. L. Richards, J. Appl. Phys. \textbf{76} 1 (1994).

\bibitem{Kadin1990} A. M. Kadin, M. Leung and A. D. Smith, Phys. Rev. Lett. \textbf{65}, 25 (1990).

\bibitem{Irwin1996} K. D. Irwin, G. C. Hilton, D. A. Wollman, and J. M. Martinis, Appl. Phys. Lett. \textbf{69} 13 (1996).

\bibitem{Baselmans2012} J. J. A. Baselmans, J. Low Temp. Phys. \textbf{167}, 292 (2012)

\bibitem{Goltsman2001} G. N. Gol'tsman, O. Okunev, G. Chulkova, A. Lipatov, A. Semenov, K. Smirnov, B. Voronov, A. Dzardanov, C. Williams and R. Sobolewski, Appl. Phys. Lett. \textbf {79}, 705 (2001).

\bibitem{Hadfield2009} R. H. Hadfield, Nat. Photonics \textbf {3}, 696 (2009).

\bibitem{Takesue2007} H. Takesue, S. W. Nam, Q. Zhang, R. H. Hadfield, T. Honjo, K. Tamaki abd Y. Yamamoto, Nat. Photonics \textbf{1} 343 (2007)

\bibitem{Mccarthy2013} A. McCarthy, N. J. Krichel, N. R. Gemmell, X. Ren, M. G. Tanner,S. N. Dorenbos, V. Zwiller, R. H. Hadfield and G. S. Buller, Optics Express \textbf{21} 7 (2013).

\bibitem{Marsili2013} F. Marsili, V.B .Verma, J. A. Stern, S. Harrington, A. E. Lita, T. Gerrits, I. Vayshenker, B. Baek, M. D. Shaw, R. P. Mirin and S. W. Nam, Nat Photonics \textbf{7} 210 (2013)

\bibitem{Natarajan2012} C. M. Natarajan, M. G Tanner and R. H Hadfield, Sup. Sci. and Technology \textbf{25} 063001 (2012)

\bibitem{Lusche2014} R. Lusche, A. Semenov, K. Ilin, M. Siegel, Y. Korneeva, A. Trifonov, A. Korneev, G. Goltsman, D. Vodolazov and H. W. Hübers, J. Appl. Phys. \textbf {116}, 043906 (2014).

\bibitem{Korneeva2018} Y. Korneeva, D. Yu. Vodolazov, A. V. Semenov, I. Florya, N. Simonov, E. Baeva, A. A. Korneev, G. N. Goltsman and T. M. Klapwijk, Phys. Rev. Applied \textbf{9} 064037 (2018)

\bibitem{RevoJPCS1965} E. Revolinsky, G. A. Spiering, D. J. Beerntsen, J. Phys. Chem. Solids \textbf{26}, 1029 (1965)

\bibitem{FrinPRL1972} R. F. Frindt, Phys. Rev. Lett. \textbf{28}, 299 (1972).

\bibitem{Cao2015} Y. Cao, A. Mishchenko, G. L. Yu, E. Khestanova, A. P. Rooney, E. Prestat, A. V. Kretinin, P. Blake, M. B. Shalom, C. Woods, J. Chapman, G. Balakrishnan, I. V. Grigorieva, K. S. Novoselov, B. A. Piot, M. Potemski, K. Watanabe, T. Taniguchi, S. J. Haigh, A. K. Geim \emph{et al.}, Nano Lett. \textbf {15}, 4914 (2015).

\bibitem{Yan2012} J. Yan, M-H.Kim, J. A. Elle, A. B. Sushkov, G. S. Jenkins, H. M. Milchberg, M. S. Fuhrer and H. D. Drew, Nat. Nanotechnology \textbf {7} 472 (2012)

\bibitem{TL} https://www.thorlabs.com/

\bibitem{Donati} S. Donati \emph{Photodetectors: Devices, Circuits and Applications}, (Prentice Hall, 2000).

\bibitem{CasiNL2007} C. Casiraghi, A. Hartschuh, E. Lidorikis, H. Qian, H. Harutyunyan, T. Gokus, K. S. Novoselov and A. C. Ferrari, Nano Lett. \textbf {7}, 2711 (2007).

\bibitem{BenaN2011} M. M. Benameur, B. Radisavljevic, J. S. Heron, S. Sahoo, H. Berger and A. Kis, Nanotechnology \textbf {22}, 125706 (2011).

\bibitem{HQG} http://www.hqgraphene.com/

\bibitem{ElbaSST2013} M. S. El-Bana, D. Wolverson, S. Russo, G. Balakrishnan, D. M. Paul and S. J. Bending, Supercond. Sci. Tech. \textbf {26}, 125020 (2013).

\bibitem{TaniJCG2007} T. Taniguchi and K. Watanabe, J. Cryst. Growth \textbf{303}, 525 (2007).

\bibitem{NovoPNAS2005} K. S. Novoselov, D. Jiang, F. Schedin, T. J. Booth, V. V. Khotkevich, S. V. Morozov and A. K. Geim, Proc. Natl Acad. Sci. USA \textbf{102}, 10451 (2005).

\bibitem{Cast2DM2014} A. Castellanos-Gomez, M. Buscema, R. Molenaar, V. Singh, L. Janssen, H. S. J. van der Zant and G. A. Steele, 2D Mater. \textbf {1}, 011002 (2014).

\bibitem{PalaNC2016} C. Palacios-Berraquero, M. Barbone, D. M. Kara, X. Chen, I. Goykhman, D. Yoon, A. K. Ott, J. Beitner, K. Watanabe, T. Taniguchi, A. C. Ferrari and M. Atatüre, Nat. Commun. \textbf {7}, 12978 (2016).

\bibitem{PurdNC2018} D. G. Purdie, N. M. Pugno, T. Taniguchi, K. Watanabe, A. C. Ferrari and A. Lombardo, Nat. Commun. \textbf {9}, 5387 (2018).

\bibitem{FerrNN2013} A. C. Ferrari and D. M. Basko, Nat. Nanotechnology \textbf {8}, 235 (2013).

\bibitem{ZhanPRB2013} X. Zhang, W. P. Han, J. B. Wu, S. Milana, Y. Lu, Q. Q. Li, A. C. Ferrari and P. H. Tan, Phys. Rev. B \textbf {87}, 115413 (2013).

\bibitem{TanNM2012} P. H. Tan, W. P. Han, W. J. Zhao, Z. H. Wu, K. Chang, H. Wang, Y. F. Wang, N. Bonini, N. Marzari, N. Pugno, G. Savini, A. Lombardo and A. C. Ferrari, Nat. Mater. \textbf {11}, 294 (2012).

\bibitem{ReicPRB2005} S. Reich, A. C. Ferrari, R. Arenal, A. Loiseau, I. Bello and J. Robertson
Phys. Rev. B \textbf {71}, 205201 (2005).

\bibitem{ArenNL2006} R. Arenal, A. C. Ferrari, S. Reich, L. Wirtz, J.-Y. Mevellec, S. Lefrant, A. Rubio and A. Loiseau, Nano Lett. \textbf {6}, 1812 (2006).

\bibitem{NemaPRB1981} R. J. Nemanich, S. A. Solin, and R. M. Martin, Phys. Rev. B \textbf {23}, 6348 (1981).

\bibitem{Li2016} J. Li, R. A. Kirkwood, L. J. Baker, D. Bosworth, K. Erotokritou, A. Banerjee, R. M. Heath, C. M. Natarajan, Z. H. Barber, M. Sorel, and R. H. Hadfield, Opt. Express \textbf{24}, 13931 (2016)

\bibitem{Tinkham2004Introduction} M. Tinkham, \emph{Introduction to Superconductivity: Second Edition}, (Dover Publications, 2004).

\bibitem{Zotova2013} A. N. Zotova and D. Y. Vodolazov, Supercond. Sci. Technol. \textbf {26},  075008 (2013).

\bibitem{Skocpol1974} W. J. Skocpol, M. R. Beasley and M. Tinkham, J. Appl. Phys. \textbf {45}, 4054 (1974).

\bibitem{Blois2017} A. Blois, S. Rozhko, L. Hao, J. C. Gallop and E. J. Romans, Supercond. Sci. Technol. \textbf {30}, 014003 (2017).

\bibitem {Zotova2012} A. N. Zotova and D. Y. Vodolazov, Phys. Rev. B \textbf {85}, 024509 (2012).

\bibitem{Boaknin2003} E. Boaknin, M. A. Tanatar, J. Paglione, D. Hawthorn, F. Ronning, R. W. Hill, M. Sutherland, L. Taillefer, J. Sonier, S. M. Hayden, and J. W. Brill, Phys. Rev. Lett. \textbf{90}, 117003 (2003).

\end{thebibliography}
\end{document}